\documentclass[preprint,showpacs,prl,aps]{revtex4}
\usepackage[T2A]{fontenc}
\usepackage[cp1251]{inputenc}
\usepackage[english]{babel}
\usepackage{latexsym}
\usepackage{indentfirst}
\usepackage{amssymb}
\usepackage{epsfig}
\usepackage{graphicx}

\begin{document}

\title{Rubidium lifetime in a dark magneto-optical trap}

\author{O.I.Permyakova, A.V.Yakovlev,
P.L.Chapovsky}

\affiliation{Institute of Automation and Electrometry,
          Russian Academy of Sciences, 630090 Novosibirsk, Russia}

\date{\today}

\vspace{1cm}
\begin{abstract}
Measurements of rubidium lifetime in a dark magneto-optical trap
(DMOT) was performed at various populations of the bright and dark
hyperfine states of the trapped atoms. The rubidium lifetime in
the trap appeared to be shorter if the atom spent more time in the
bright state. A simple explanation of this effect is based on the
increase of the cross-section of rubidium collisions with the surrounding
warm atoms upon rubidium electronic excitation. \vspace{1cm}
\end{abstract}

\vspace{2cm} \pacs{32.80.Pj}

\email{chapovsky@iae.nsk.su}

\maketitle

\section{Introduction}

Laser cooling of neutral atoms in magneto-optical traps, MOT
\cite{Raab87PRL,Monroe90PRL} became an important technique of the
modern atomic physics. MOT is a convenient and reliable apparatus
for cooling large clouds of atoms to the temperatures in the 10 -
100 $\mu$K range. The main applications of MOTs include laser
spectroscopy of cold atoms, cold collisions, new frequency
standards, first stage of cooling to achieve BEC and many others.

There are well-known difficulties in straitforward applications of
MOTs to the investigations of cold atoms. The MOTs use strong
laser radiation and inhomogeneous magnetic field that disturb
captured atoms. One way to solve this problem is to separate in
time the capture and investigation of cold atoms. A disadvantage
of this approach is that one has to work with unstable object, the
atomic cloud released from the trap.

In some cases the problem can be solved with the help of a
specially modified MOT, the dark magneto-optical trap (DMOT)
proposed in \cite{Ketterle93PRL} for alkali atoms. The DMOT
consists of two regions. The outer region has both the trapping
and repumping light and works similar to an ordinary MOT. In the
inner region the repumping light is blocked and atoms are
collected on the "dark"\ hyperfine state noninteracting with the
trapping light. Because of the scattering of the repumping light
by the optical elements and warm atoms a low intensity repumping
light is nevertheless present in the inner region of DMOT.
Consequently, some fraction of the trapped atoms is situated in
the "bright"\ hyperfine state that interacts with the trapping
light. Note, that the atoms in the "bright"\ state are responsible
for the cloud fluorescence in DMOT.

Main application of DMOTs is for increasing the density of trapped
atoms. It occurs because the particle repulsion induced by the
radiation exchange and the rate of superelastic collisions are
suppressed in DMOT. In addition to this useful ability of DMOT to
hold large fraction of atoms in dark state, DMOT allows us to vary
in a controllable way the populations of the bright and dark
states. The populations of these states determine a number of
processes in DMOT, e.g., the laser power absorbed by cold atoms.
Thus the change of the fraction of "bright"\ atoms in DMOT can be
used to study the physics of cold atoms. In this paper we report
on an experimental observation of the lifetime dependence of
rubidium atoms in a DMOT on the relative abundance of the bright
and dark hyperfine states.

\section{Rubidium level scheme and DMOT}

\begin{figure}[htb]
\includegraphics[height=12cm]{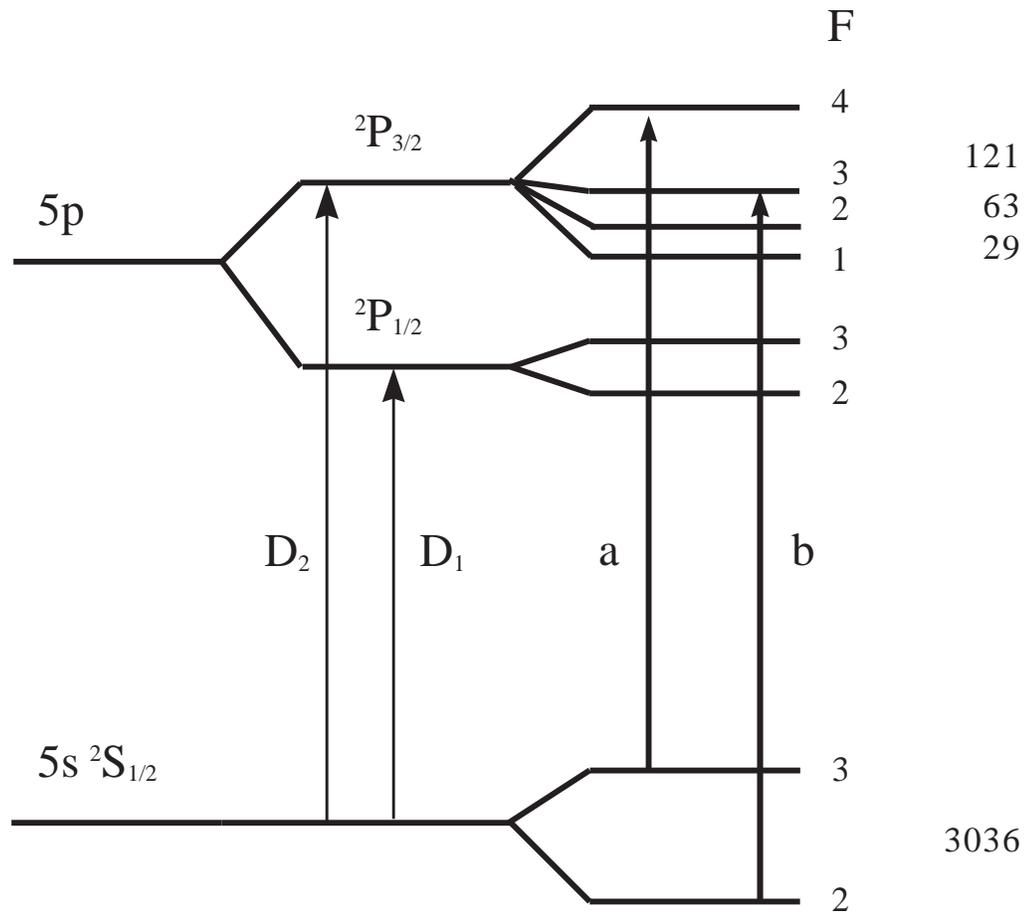}
\vspace{1cm} \caption{\sl Energy level diagram of $^{85}$Rb. a -
trapping radiation; b - repumping radiation. Numbers at the right
side give the splittings of the hyperfine levels
\cite{Barwood91APB}.} \label{fig1}
\end{figure}

Our experiment was done with $^{85}$Rb atoms. The energy level
diagram is presented in Fig.{\ref{fig1}}. The DMOT used two
lasers. The trapping laser frequency was tuned to the red by 15 -
20 MHz from the transition $F_g=3\rightarrow F_e=4$. The frequency
of the repumping laser was varied in a wide range in the vicinity
of the transitions $F_g=2\rightarrow F_e=2,3$ (see below). The
hyperfine states $F_g=2$ and $F_g=3$ are the dark and bright
states, respectively.

\begin{figure}[htb]
\includegraphics[height=10cm]{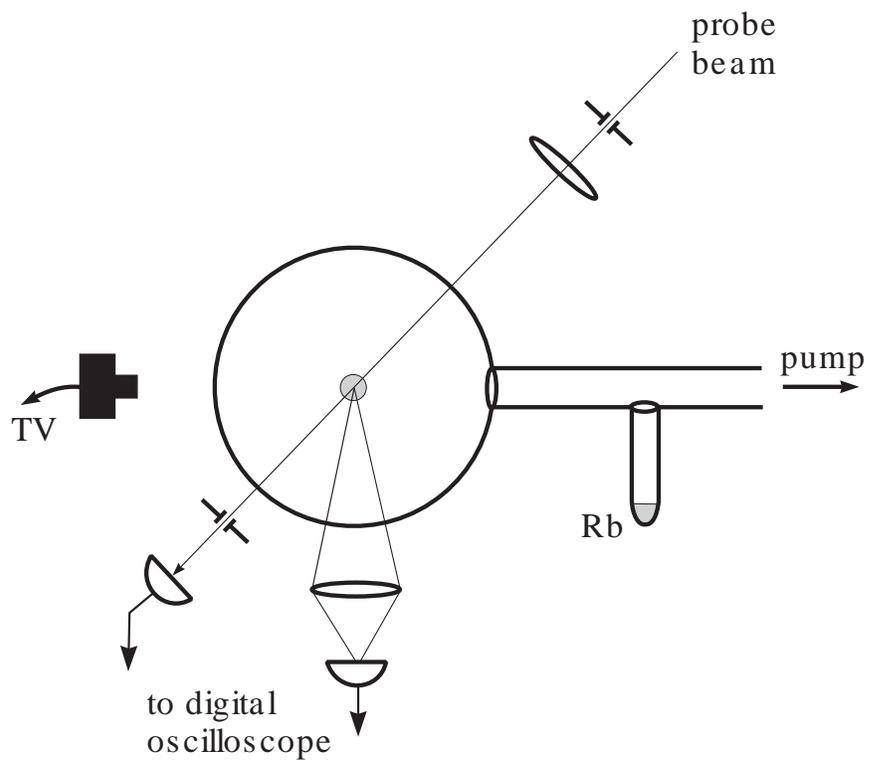}
\vspace{1cm} \caption{\sl Scheme of the measurements of the
density of trapped rubidium atoms. The Helmholtz coils, trapping
and repumping beams are not shown.} \label{fig2}
\end{figure}

An experimental setup is shown in Fig.\ref{fig2}. Our DMOT
collects atoms from surrounding warm rubidium vapour similar to
the approach suggested in the Ref.~\cite{Monroe90PRL}. The DMOT
uses 6 circular polarized trapping beams and 2 linear polarized
repumping beams all having diameter of approximately 15 mm and
intensity 6 mW. The hollow repumping beams (inner diameter equal 5
mm) were directed to the trap center at 90$^0$ to each other. The
trapping and repumping beams were produced by the two commercial
lasers DL100 (Toptica).

Quadrupole magnetic field was created by two Helmholtz coils
having 100 turns each, diameter equal 5 cm and separated by 5 cm.
The coils produced in the trap center the magnetic field gradient
equal 17~G/cm along the symmetry axis at the current through the
coils equal 1.6 A. The quadrupole magnetic field could be switched
on and off as fast as 500 $\mu$sec.

We have used simple vacuum chamber for DMOT that is a glass sphere
of diameter 6 cm  without antireflection coating. The chamber was
permanently connected to the 5 liter/sec ion pump and to the glass
appendix containing metal rubidium. The rubidium vapour pressure
in the chamber was regulated by cooling this appendix with the
help of Peltier element. Because the chamber was permanently
pumped by the ion pump the rubidium vapour pressure was
significantly lower than the saturated vapour pressure for a given
appendix temperature. The pressure of residual gases in the
chamber was on the order of 10$^{-8}$~Torr.

The rubidium populations in the ground hyperfine states were
measured in the experiment with the help of an additional probe
beam that was produced by the free running (without an external
resonator) home-made diode laser equipped with good quality
current and temperature stabilizers. The diode ML6XX24
(Mitsubishi) was used in this laser. The linewidth of the probe
radiation was estimated being less than 1 MHz (HWHM).

All three lasers used in the present experiment were equipped with
the home-made frequency stabilization systems based on the DAVLL
(dichroic atomic vapour laser lock) principle
\cite{Corwin98AO,Yashchuk00RSI}. This stabilization system allowed
us to stabilize the radiation frequency at any point in a wide
range ($\simeq$1 GHz) near each hyperfine Rb absorption lines. The
system provided a good frequency stability with the laser
frequency drift being less than 3 MHz/hour. More details on our
stabilization system can be found in \cite{Permyakova05QE}.

\section{Experimental results}

The steady-state number of captured rubidium atoms depends on the
density of surrounding warm rubidium atoms. If the appendix
containing metal rubidium was kept at room temperature, the total
number of captured rubidium atoms in our DMOT exceeded
$5\cdot10^8$. The cloud had almost spherical shape with the
diameter 0.8 mm. The shape of the rubidium cloud was determined
from the digitized fluorescence image recorded by the CCD video
camera.

\begin{figure}[htb]
\includegraphics[height=16cm]{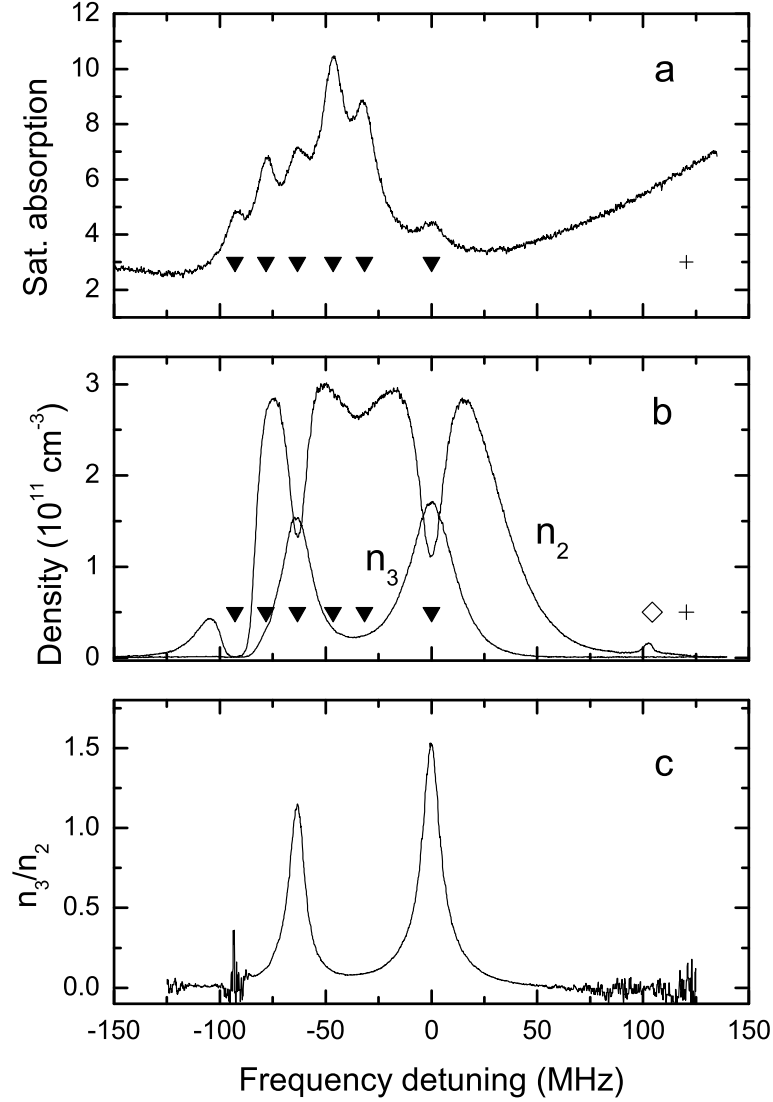}
\vspace{1cm} \caption{\sl Densities of rubidium atoms in the
hyperfine states $F_g=2$ and $F_g=3$ as a function of repumping
frequency. a - saturated resonances for the transitions
$F_g=2\rightarrow F_e=1,2,3$; b - densities in the states $F_g=2,\
(n_2)$ and $F_g=3,\ (n_3)$; c - ratio $n_3/n_2$. The signs
indicate: $\blacktriangledown$ - the positions of the saturated
resonances \cite{Barwood91APB}; {\bf +} - center of the forbidden
transition $F_g=2\rightarrow F_e=4$; $\lozenge$ - position of the
Raman resonance for the trapping and repumping radiations.}
\label{fig3}
\end{figure}

The density of captured rubidium atoms as a function of frequency
of repumping radiation is presented in Fig.~\ref{fig3}. These data
were obtained with the rubidium appendix kept at room temperature.
In order to determine the value of the repumping frequency the
saturation absorption resonances of counter propagating beams in
an additional rubidium cell (at room temperature) were recorded.
These resonances for the hyperfine transitions $F_g=2\rightarrow
F_e=1,2,3$ are shown in Fig.~\ref{fig3}a. Because there are three
allowed transitions in the absorption spectra $F_g=2\rightarrow
F_e=1,2,3$, the saturation resonance spectrum has six peaks.
Frequency calibration of this spectrum was performed using the
tabulated positions of the resonances given in
\cite{Barwood91APB}. For convenience of the reader the frequencies
of the saturated resonances are listed in Table~I.

The Rb densities in the $F_g=2$ state ($n_2$) and in the $F_g=3$
state ($n_3$) at various  repumping frequencies are presented in
Fig.~\ref{fig3}b. These data were recorded at slow scan rate of
the  repumping frequency in order to account the large filling
time of the trap. The $n_3$ curve has two peaks at the repumping
frequencies coinciding with the transitions $F_g=2\rightarrow
F_e=2,3$. The $n_2$ curve has more complicated structure. For the
following we need to consider in the $n_2$ curve the two positive
wide peaks and the two negative narrow peaks all centered at the
repumping frequencies of $F_g=2\rightarrow F_e=2,3$ transitions.
The positive wide peaks in the $n_2$ and $n_3$ curves are due to
the peaks in the efficiency of the repumping radiation in the
outer region of DMOT. The narrow negative peaks in the $n_2$ curve
are due to the hyperfine pumping caused by the weak scattered
repumping light in the inner region of DMOT. More details of this
experiment and the discussion of the curves' features can be found
in \cite{Chap06JETP}.

Let us consider now the ratio of the rubidium densities in $F_g=3$
and $F_g=2$ states, $n_3/n_2$, as a function of repumping
frequency (Fig.~\ref{fig3}c). The ratio $n_3/n_2$ has two maxima
at the centers of $F_g=2\rightarrow F_e=2,3$ transitions.  The
ratio $n_3/n_2$ has deep minimum at the repumping frequency
between the $F_g=2\rightarrow F_e=2,3$ transitions. Note that the
DMOT capture remains highly efficient at all these repumping
frequencies. The data in Fig.~\ref{fig3}c shows that by tuning the
repumping frequency the ratio $n_3/n_2$ in the DMOT can be varied
from 1.5 to 0.08, thus by 20 times.

These features of our DMOT was used in the present experiment to
measure the state dependence of the rubidium lifetime in the trap.
We measured the number of trapped atoms in the $F_g=2$ state as a
function of time after switching on the quadrupole magnetic field
at the three positions of repumping frequency, $\omega_{rep}$:
\begin{equation}\label{omega}
\omega_{rep}=\omega_{22},\ \ \omega_{rep}=\omega_{23},\ \
\omega_{rep}=-40~\text{MHz},
\end{equation}
where $\omega_{22}$ and $\omega_{23}$ are the frequencies of the
$F_g=2\rightarrow F_e=2,3$ transitions, respectively. The number
of trapped atoms was measured with the help of the probe beam
absorption. The probe frequency was tuned to the center of the
$F_g=2\rightarrow F_e=1$ transition. This transition is a closed
transition and thus experiences smaller power saturation that the
two other allowed transitions started from the state $F_g=2$.

The measurements were done at rubidium appendix temperature equal
-8~C. This reduced significantly the rubidium pressure in the trap
chamber and consequently the number of trapped atoms became 10 -
20 times smaller than for rubidium appendix being at the room
temperature. The decrease of the number of trapped atoms
simplified the experimental situation because the cloud appeared
to be optically thin. On the other hand, it was checked that the
ratio $n_2/n_3$ undergoes the similar variations with the change
of $\omega_{rep}$ as in the Fig.~\ref{fig3}c measured for the
rubidium appendix at the room temperature.

\begin{figure}[htb]
\includegraphics[height=12cm]{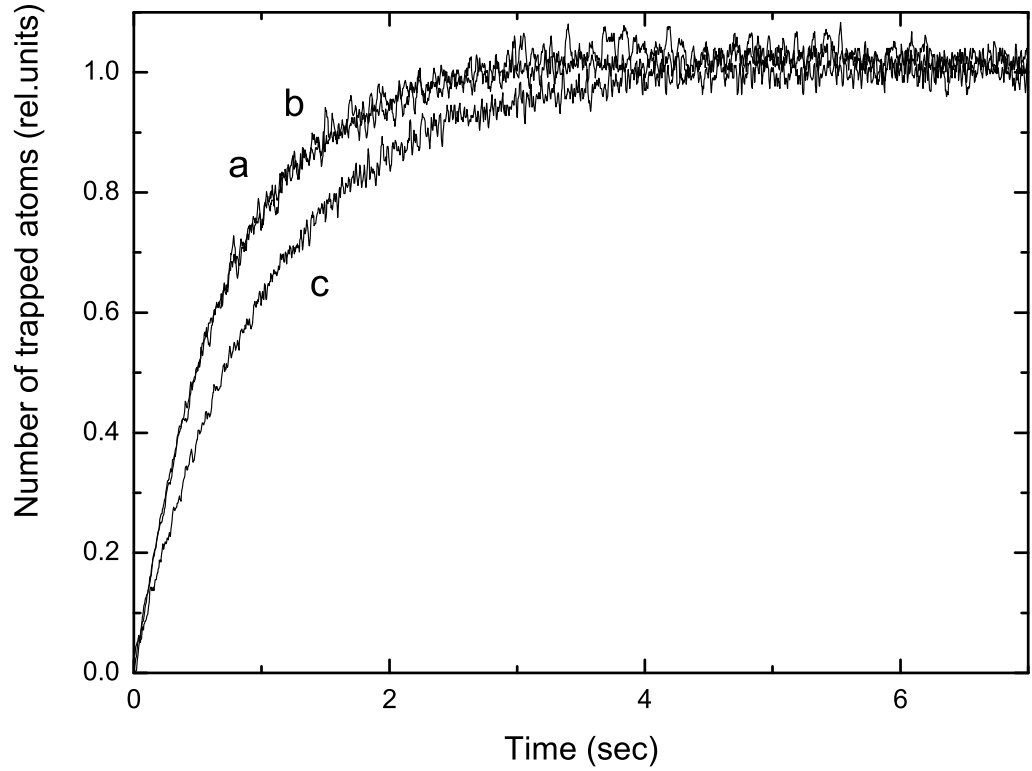}
\vspace{1cm} \caption{\sl The time dependences of the number of
captured rubidium atoms for the three positions of the repumping
frequency, $\omega_{rep}$. Curve a: $\omega_{rep}$ is at the
center of the $F_g=2\rightarrow F_e=2$ transition; curve b:
$\omega_{rep}$ is at the center of the $F_g=2\rightarrow F_e=3$
transition; curve c: $\omega_{rep}=-40$~MHz.} \label{fig4}
\end{figure}

The time dependencies of the number of captured atoms are
presented in Fig.~\ref{fig4}.   The two curves ("a" and "b")
corresponding to $\omega_{rep}$ being at the centers of the
$F_g=2\rightarrow F_e=2,3$ transitions coincide completely with
each other. The third curve for $\omega_{rep}=-40$~MHz shows
significantly longer lifetime. The measured lifetimes are,
\begin{equation}\label{tau}
\tau_{22}=0.72\pm0.01~\text{sec};\ \
\tau_{23}=0.74\pm0.01~\text{sec};\ \
\tau_{mid}=1.03\pm0.01~\text{sec};\ \
\end{equation}
where indices refer to the position of the repumping frequency.
The filling time dependencies presented in Fig.~\ref{fig4}
appeared to be purely exponential at the experimental accuracy.

\section{Discussion}

The lifetime of rubidium atoms in magneto-optical traps depends on
many effects (see, e.g., the review \cite{Weiner99RMP}). One can
neglect some of these effects at low density of captured atoms
chosen in this experiment for the rubidium lifetime measurements.
The fact that our filling curves were purely exponential supports
the neglect of the reabsorption of rubidium fluorescence and
superelastic collisions inside the cloud. In this simplified case,
the number of trapped atoms, $N$, is governed by the following
equation \cite{Monroe90PRL},
\begin{equation}\label{fill}
    \frac{dN}{dt}=R -N\gamma,
\end{equation}
where, $R$ is the trap capture rate of rubidium atoms and $\gamma$
is the escape rate of rubidium atoms from the trap. The rate
$\gamma$ is determined by the collisions with the surrounding warm
atoms,
\begin{equation}\label{gamma}
\gamma=n_{Rb}\sigma_{Rb}v_{Rb} + n_b\sigma_bv_b,
\end{equation}
where $n_{Rb}$, $n_b$, are the concentrations of warm rubidium
atoms and residual gases, $v_{Rb}$, $v_b$ are their velocities.
The cross-sections, $\sigma_{Rb}$ and $\sigma_b$, characterize the
collisions that accelerate cold atoms to the velocity larger than
the critical velocity, $v_c$. Critical velocity is the maximum
velocity of Rb atoms that trap is able to capture
\cite{Monroe90PRL}.

The model of Eq.~(\ref{fill}) gives the exponential time
dependence for the number of captured atoms with the time constant
equal $\gamma^{-1}$. The observed in our experiment change in the
rubidium lifetime at various repumping frequencies can be
understood if one suggests that the cross-sections $\sigma_{Rb}$
and $\sigma_b$ depend on the electronic state of cold rubidium
atom. Note that the times spent in the ground and in the
electronically excited states of the trapped rubidium atoms differ
significantly for the repumping frequencies equal $\omega_{22}$,
or $\omega_{23}$ and $\omega_{mid}$ chosen in this experiment. The
measurements made above show that there are 60\% of Rb atoms in
the bright hyperfine state if $\omega_{rep}$ equals to
$\omega_{22}$, or $\omega_{23}$, and only 10\% if
$\omega_{rep}=-40$~MHz. Rb atoms in the bright state spend
approximately half of their time in the excited state.
Consequently, the fraction of the electronically excited Rb atoms
equals $\simeq$30\% if $\omega_{rep}=\omega_{22}$, or
$\omega_{23}$ and $\simeq$5\% if $\omega_{rep}=-40$~MHz. One can
estimate that such variation of electronically excited Rb atoms
can explain the change of lifetime given by Eq.(\ref{tau}) if an
electronically excited Rb atom has its "escape"\ cross-section 3
times larger than an unexcited atom.

The state dependence of the rubidium cross-sections is the fact
well established in the experiments devoted to the rubidium
light-induced drift (LID) effect \cite{Wittgrefe90b}. It was
measured in Ref.~\cite{Wittgrefe90b} that the change in the
rubidium transport cross-section can be as as large as 50\%. Note
also, that the losses in magneto-optical traps are dependent on
similar but not identical cross-sections. The LID effect depends
on the transport cross-section change upon atom excitation, but
the trap losses are dependent on the cross-section of the transfer
to the cold atom a small amount of kinetic energy sufficient for
the atom to escape from the trap. This "escape"\ energy is by 2-3
orders of magnitude smaller than the $k_BT$ at room temperature.

\section{Conclusions}

We have measured the rubidium lifetime in the dark magneto-optical
trap (DMOT) at the relative populations of the bright and dark
hyperfine states different by factor 20. The measurements showed
that the rubidium lifetime in DMOT became shorter when it spent
more time in the bright state. The possible explanation of this
effect can be based on the assumption that the cross-section of
trapped rubidium atoms with surrounding warm atoms is larger for
the electronically excited rubidium than for rubidium in the
ground state. Similar trend was observed previously for the
transport cross-section of electronically excited rubidium atoms
in the investigations of the light-induced drift effect (LID)
\cite{Wittgrefe90b}.

\section{Acknowledgments}

The work was supported by the Russian Foundation for Basic
Research (grants RFBR 06-02-16415; 06-02-08134 ), the Siberian
Division and the Division of General Physics and Astronomy of the
Russian Academy of Sciences.


\vspace{2cm}
\begin{table}[htb]
\centering
\begin{tabular}{|c|c|c|c|c|c|c|}\hline
  Frequency detuning (MHz) & -92.82         & -78.11         & -63.4         & -46.41         & -31.7         & 0 \\\hline
  Transition       & $2\rightarrow1$ & $2\rightarrow1$ & $2\rightarrow2$ & $2\rightarrow1$ & $2\rightarrow2$ & $2\rightarrow3$ \\
                &                 & $2\rightarrow2$ &                 & $2\rightarrow3$ & $2\rightarrow3$ &  \\\hline
\end{tabular}
\caption{Frequencies of the saturated absorption resonances in
$^{85}$Rb for the transitions started from the $F_g=2$ hyperfine
state \cite{Barwood91APB}. The double transitions indicate the
cross-resonances. Zero frequency detuning was chosen at the center
of the F$_g$=2->F$_e=3$ transition.}\label{table2}
\end{table}

\newpage

\newpage

\newpage

\end{document}